\documentclass{article}
\usepackage{amsmath}
\usepackage{amssymb}
\usepackage{graphicx}
\usepackage{cite}
\usepackage{xcolor}

\newcommand{\new}[1]{#1}

\newcommand{\e}{\varepsilon}              %
\newcommand{\vp}{{\varphi}}              %
\newcommand{\w}{{\omega}} 
\newcommand{\W}{{\Omega}}

\begin{document}

\title{Internal Reliability of Coupled Kuramoto-Sakaguchi Phase Oscillators}

\author{
Arkady Pikovsky$^1$, Franco Bagnoli$^{2,3}$ , Stefano Iubini$^{4,3}$
\\[0.5cm]
\begin{minipage}{\textwidth}
$^1$ Institute for Physics and Astronomy, University of Potsdam, Potsdam, Germany.
\texttt{pikovsky@uni-potsdam.de}\\
$^2$ Department of Physics and Astronomy and CSDC, University of Florence, via G. Sansone 1, I-50019, Sesto Fiorentino, Italy. \texttt{franco.bagnoli@unifi.it}\\
$^3$ Istituto Nazionale di Fisica Nucleare, Sezione di Firenze, via G. Sansone 1, I-50019, Sesto Fiorentino, Italy\\
$^4$ Istituto dei Sistemi Complessi, Consiglio Nazionale delle Ricerche, via Madonna del Piano 10, I-50019, Sesto Fiorentino, Italy
\end{minipage}
}

\maketitle

\begin{abstract}
The notion of internal reliability in dynamical networks describes whether replicas of a particular unit follow the dynamics of the reference unit. Reliability and anti-reliability can be quantified by the transversal Lyapunov exponents. We study phase oscillators coupled via Kuramoto-Sakaguchi-type interactions. Already the simplest solvable system of two oscillators demonstrates nontrivial reliability properties. We present numerical evidence of reliability and anti-reliability in small networks with a uniform distribution of natural frequencies. The dynamics of an ensemble of replicas can be described within the Watanabe-Strogatz theory, which predicts symmetry of the transversal Lyapunov exponents for replica-attractor and replica-repeller.  \\
\textbf{keywords:} {Kuramoto-Sakaguchi model, reliability, transversal Lyapunov exponent, synchronization}
\end{abstract}

\section{Introduction}
\label{sec:intro}

Synchronization by common noise means that the states of two identical nonlinear systems driven by the same noise converge and eventually coincide despite different initial conditions~\cite{Pikovsky-84a}. In neurosciences, this effect is called reliability~\cite{Mainen-Sejnowski-95,ermentrout2008reliability,teramae2007reliability,lin2009reliability}.
In the reliability setup, one has a single system (a neuron) and applies the same pre-recorded noise signal to it repeatedly, assuming the system's properties remain the same (apart from initial conditions). If repeated driving of a system with the same noise force produces the same response, one calls such a situation reliable, which corresponds to synchronization. Reliability and synchronization by common noise are quantified by the largest Lyapunov exponent of the noise-driven system: a negative Lyapunov exponent means synchronization and reliability, while a positive Lyapunov exponent means sensitivity to initial conditions, desynchronization, and anti-reliability~\cite{Goldobin-Pikovsky-06b}. Remarkably, the same concept under the name of damage spreading has been introduced in the field of Monte-Carlo simulations of 
models with discrete state space, like the Ising model or probabilistic cellular automata~\cite{stanley1987dynamics,herrmann1990damage,grassberger1995damage}. Here, common noise means that the same sequence of random numbers is used in the Monte-Carlo steps for two systems that differ only by an initial state. Synchronization (reliability) in this setup is called ``healing'' (damage disappears), while the absence of synchrony (anti-reliability) is referred to as a chaotic situation. Although the usual notion of the Lyapunov exponent is not applicable for systems with discrete states, one can use Boolean derivatives to properly define the Lyapunov exponent in the realm of damage spreading~\cite{bagnoli1999synchronization}.

\begin{figure}
    \centering
    \includegraphics[width=0.5\linewidth]{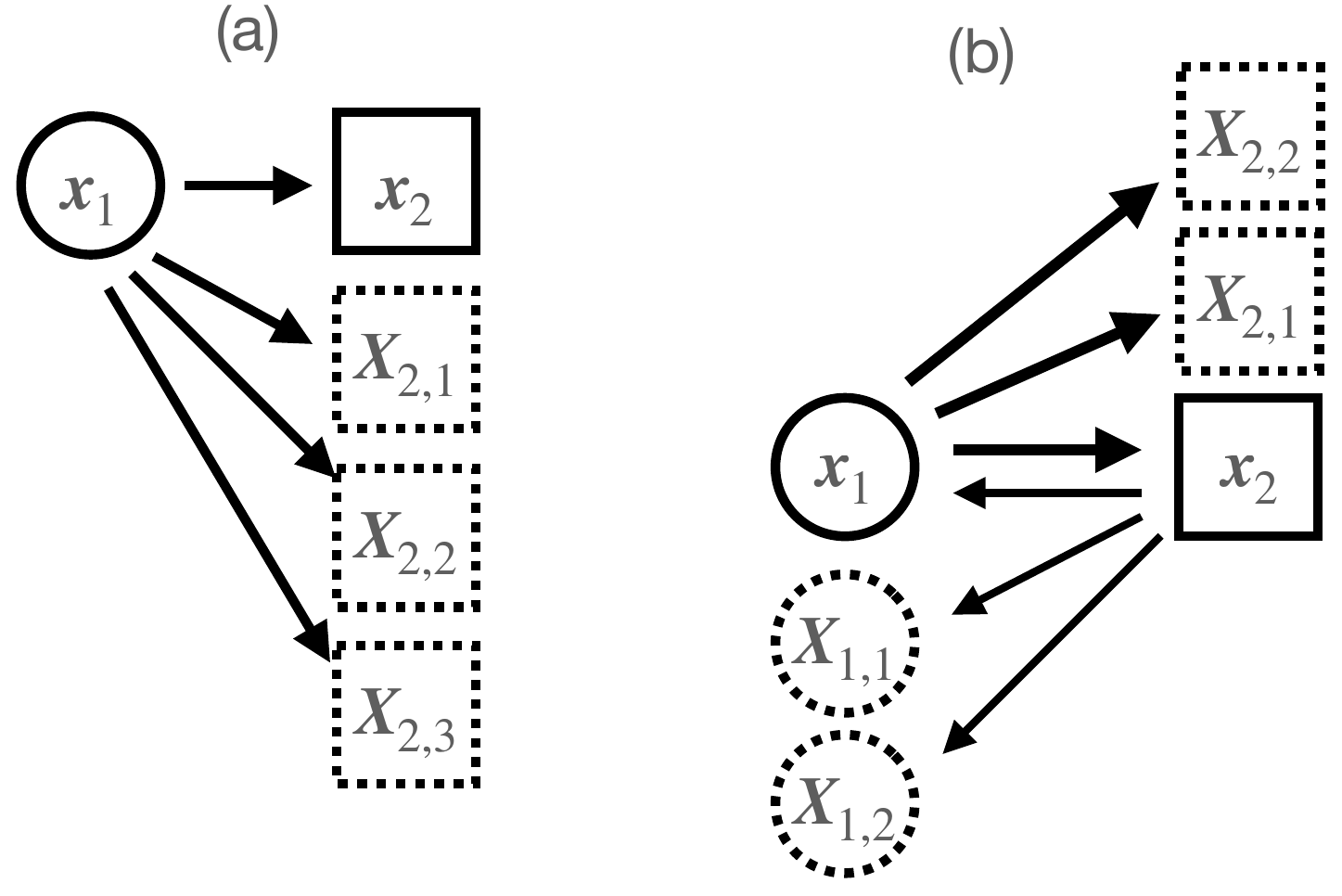}
    \caption{\new{ (a) A sketch of the generalized synchronization problem;  (b) a sketch of the internal reliability one. In the generalized synchronization, there is a unidirectional coupling $1\to 2$, while in the internal reliability setup the coupling is bidirectional $1\leftrightarrow 2$. The driving force of replicas ($\boldsymbol{X}_{k,m}$) is the same as that of the prototype ($\boldsymbol{x}_k$).}}
    \label{fig:sk}
\end{figure}

If one replaces noisy driving by a chaotic one, resulting from a chaotic dynamical system, then synchronization by common noise and reliability coincide with the notion of generalized synchronization of chaos~\cite{Rulkov-Sushchik-Tsimring-Abarbanel-95,abarbanel1996generalized,Baia2025,Letelier-23,letellier2024taxonomy}. 
\new{Generalized synchronization setup is illustrated in Fig.~\ref{fig:sk}(a), where the driving unit $\boldsymbol{x}_1$ not only drives the unit $\boldsymbol{x}_2$, but also its three replicas $\boldsymbol{X}_{2,1},\boldsymbol{X}_{2,2},\boldsymbol{X}_{2,3}$. Synchronization occurs if at large times $\boldsymbol{x}_2=\boldsymbol{X}_{2,1}=\boldsymbol{X}_{2,2}=\boldsymbol{X}_{2,3}$}

In a recent paper \cite{Matteuzzi2025}, we extended the notion of reliability to the case where different units are coupled bi-directionally (and not uni-directionally, as in the generalized synchronization setup, \new{see Fig.~\ref{fig:sk}}. One chooses one subunit of a complex network (we call this unit a prototype) and prepares one or several identical replicas of this unit. These replicas receive the same input from the rest of the network as the prototype receives, but they do not influence other units and remain ``passive''. The replicas can have different initial states. \new{As an example, we present in Fig.~\ref{fig:sk}(b) two bidirectionally interacting units $\boldsymbol{x}_1,\boldsymbol{x}_2$, and prepare two replicas of each unit. The replicas of $\boldsymbol{x}_1$ (denoted as $\boldsymbol{X}_{1,1},\boldsymbol{X}_{1,2}$) are driven by the same force from unit $\boldsymbol{x}_2$ as the unit $\boldsymbol{x}_1$.}  If the states of the replicas in course of time converge to the state of the prototype \new{(for example, $\boldsymbol{X}_{1,1}\to \boldsymbol{x}_1$,
 $\boldsymbol{X}_{1,2}\to \boldsymbol{x}_1$)}, one calls this unit reliable; otherwise, if the states of the replicas remain different from the state of the prototype, one speaks of anti-reliability.   We note here that using replicas of certain variables of a chaotic system is a central point of the Pecora-Carroll setup of chaos synchronization~\cite{Pecora1990}. The main difference of our setup in this paper is that here a network of regular oscillators is considered, and the overall dynamics can also be regular and not chaotic. 

In this paper, we discuss reliability properties of phase oscillators with a Kuramoto-Sakaguchi-type coupling~\cite{Sakaguchi-Kuramoto-86,Acebron-etal-05,chen2019dynamics,Gupta-Campa-Ruffo-18}. After presenting a general formalism of internal reliability in Section~\ref{sec:gen}, we study in detail the case of two coupled oscillators in Section~\ref{sec:two}. Here, the problem can be solved analytically. After that, in Section~\ref{sec:kse}, we discuss reliability properties of larger networks, employing numerical simulations and the Watanabe-Strogatz theory~\cite{Watanabe-Strogatz-93,Watanabe-Strogatz-94}. The results are discussed in Section~\ref{sec:concl}.

\section{General notion of internal reliability}
\label{sec:gen}
In this section, we shortly introduce the notion of internal reliability, following recent Ref.~\cite{Matteuzzi2025}. 

\subsection{Replicas and their stability}

Consider a system of $N$ coupled units $\boldsymbol{x}_j$, $j=1,\ldots,N$, described by coupled dynamical equations
\begin{equation}
\dot{\boldsymbol{x}}_j=\boldsymbol{F}_j(\boldsymbol{x}_j;\boldsymbol{x}_{l\neq j}),\quad j=1,\ldots,N.
    \label{eq:bas1}
\end{equation}
Note that all units can be completely different (i.e., they have different dimensions and different dynamics).  In this paper, we assume dissipative dynamics in Eq.~\eqref{eq:bas1}, and suppose that at $t=0$ the state  $\boldsymbol{x}_j(0)$ is on an attractor. 

Let us fix some unit $k$ and make at time $t=0$ several replicas, $\boldsymbol{X}_{k,m}(0)$, of it \new{(see sketch in Fig.~\ref{fig:sk}, where replicas are depicted with dashed lines)}. We call the replicated unit $\boldsymbol{x}_k$ the prototype. These replicas receive the same input from 
other units as $\boldsymbol{x}_k$ does, so that their dynamics is
\begin{equation}
\dot{\boldsymbol{X}}_{k,m}=\boldsymbol{F}_k(\boldsymbol{X}_{k,m};\boldsymbol{x}_{l\neq k}),\quad m=1,\ldots,M.
    \label{eq:bas2}
\end{equation}
The system described by  Eqs.~\eqref{eq:bas1} and \eqref{eq:bas2} is a skew system, where replicas are forced by the units $\boldsymbol{x}_j$, but not vice versa. It is evident that
if the initial state of a replica coincides with its prototype, i.e., $\boldsymbol{X}_{k,m}(0)=\boldsymbol{x}_k(0)$, then for all $t>0$ their states also coincide $\boldsymbol{X}_{k,m}(t>0)=\boldsymbol{x}_k(t>0)$. However, if a replica does not coincide with the prototype initially, \new{i.e., $\boldsymbol{X}_{k,m}(0)\neq\boldsymbol{x}_k(0)$}, different scenarios are possible.

If all or almost all replicas converge to the prototype $|\boldsymbol{X}_{k,m}(t)-\boldsymbol{x}_k(t)|\to 0$ as $t\to\infty$, we call the unit $k$ \textbf{reliable}. If all or almost all replicas do not converge to the corresponding prototype, we call such a unit \textbf{anti-reliable}. 

In general, there are more possibilities related to ``multi-stability": one can imagine situations where some replicas \new{of the same prototype (i.e., replicas that have the same index $k$ but different index $m$)} converge to the corresponding prototype while others do not. \new{For example, for a setup presented in Fig.~\ref{fig:sk}(b), multistability in replicas occurs if $\boldsymbol{X}_{2,1}$ converges to $\boldsymbol{x}_2$, while $\boldsymbol{X}_{2,2}$  remains at a finite distance to $\boldsymbol{x}_2$.}However, as we show below, such behavior cannot occur in phase oscillators with Kuramoto-Sakaguchi-type coupling, and we will not further elaborate on these possibilities.

It is convenient to denote a trajectory, to which replicas driven by Eq.~\eqref{eq:bas2} converge, with the term \textit{replica-attractor}. Because generally the forcing in Eq.~\eqref{eq:bas2} is irregular, this attractor is not a set in the phase space of replicas $\boldsymbol{X}$, but a trajectory to which trajectories with other initial conditions converge. A similar definition of a random attractor (sometimes called ``pullback attractor'') is used for randomly driven systems.
In the reliable case, the prototype trajectory coincides with the replica attractor.
In the anti-reliable case, the prototype trajectory is unstable. Generally, such a trajectory can be of saddle type, but for one-dimensional units like phase oscillators, there is only one unstable direction. Thus, the prototype is \textit{replica-repeller}, while the replica-attractor is a different trajectory.

To determine whether the replicas converge to the prototype or not, one has to examine their linear stability properties. \new{Let us denote the deviation of a replica $\boldsymbol{X}_k$ from the prototype $\boldsymbol{x}_k$ as $\boldsymbol{u}_k=\boldsymbol{X}_k-\boldsymbol{x}_k$. The equation for this deviation is obtained by subtracting Eq.~\eqref{eq:bas1} from Eq.~\eqref{eq:bas2}:
\[
\dot{\boldsymbol{u}}_k=\dot{\boldsymbol{X}}_k-\dot{\boldsymbol{x}}_k=\boldsymbol{F}_k(\boldsymbol{X}_{k};\boldsymbol{x}_{l\neq k})-\boldsymbol{F}_k(\boldsymbol{x}_{k};\boldsymbol{x}_{l\neq k})=\boldsymbol{F}_k(\boldsymbol{x}_{k}+\boldsymbol{u}_k;\boldsymbol{x}_{l\neq k})-\boldsymbol{F}_k(\boldsymbol{x}_{k};\boldsymbol{x}_{l\neq k})\;.
\]
Considering infinitely small deviations $\boldsymbol{u}_k$, we obtain after linearisation}
\begin{equation}
\dot{\boldsymbol{u}}_k=\frac{\partial}{\partial \boldsymbol{x}_k}\boldsymbol{F}_k(\boldsymbol{x}_k;\boldsymbol{x}_{l\neq k})\boldsymbol{u}_k\;.
    \label{eq:bas3}
\end{equation}
The linear system of Eq.~\eqref{eq:bas3}, where states $\{\boldsymbol{x}_j\}$ evolve according to Eq.~\eqref{eq:bas1}, defines transversal Lyapunov exponents~\cite{pikovsky2016lyapunov} (their number is determined by the dimension of variable $\boldsymbol{u}_k$), and it is enough to look at the largest of them, which determines the asymptotic growth rate of the perturbation $|\boldsymbol{u}_k(t)|\sim \exp(\lambda_k t)$. If the largest transversal Lyapunov exponent (LTLE) $\lambda_k$ is positive, the trajectory $\boldsymbol{x}_k(t)$ is unstable in replica phase space, and this unit is anti-reliable. If 
the LTLE is negative, then the unit is reliable. Note that the LTLE depends on the index $k$; as we will see below, there are situations where some units are reliable and some anti-reliable (it can happen that all units are reliable or all units are anti-reliable, but we never observed the latter situation for phase oscillators).

Similarly to the stability of the prototype, we can quantify the stability of the replica-attractor (qualitatively, this attractor is always stable due to the definition). Suppose that different replicas converge to an attracting trajectory $\boldsymbol{X}_k(t)$. Denoting $\boldsymbol{w}_k$ an infinitesimal difference between two replicas close to the replica-attractor, we obtain the evolution
\begin{equation}
\dot{\boldsymbol{w}}_k=\frac{\partial}{\partial \boldsymbol{X}_k}\boldsymbol{F}_k(\boldsymbol{X}_k;\boldsymbol{x}_{l\neq k})\boldsymbol{w}_k\;.
    \label{eq:bas4}
\end{equation}
Here states $\boldsymbol{X}_k,\boldsymbol{x}_{j\neq k}$ evolve according to the system of Eqs.~\eqref{eq:bas1} and \eqref{eq:bas2}. The largest Lyapunov exponent $\Lambda_{\text{attr}}$ of Eq.~\eqref{eq:bas4} determines the rate with which $\boldsymbol{w}_k$ decays: $|\boldsymbol{w}_k(t)|\sim \exp[\Lambda_{\text{attr}} t]$, with $\Lambda_{\text{attr}}<0$.

\subsection{Reliability and inference of the unit's state and parameters}

In this section, we shortly discuss the importance of reliability in a particular technique of recovering a unit's properties
from the observations. Suppose that in the system of Eq.~\eqref{eq:bas1},
we observe all the units except unit $k$. If we know the equations of motion of $\boldsymbol{x}_k$, we can try to solve  Eq.~\eqref{eq:bas2} starting from arbitrary initial conditions, in the hope that this solution $\boldsymbol{X}_k(t)$ after some transient will reproduce the true evolution of $\boldsymbol{x}_k(t)$. As it follows from the definitions of Section~\ref{sec:gen}, this is possible only if reliability holds. In an anti-reliable situation, such a method fails.

\section{Reliability of two coupled phase oscillators}
\label{sec:two}
The main topic of this paper is the exploration of the reliability properties of phase oscillators with a Kuramoto-Sakaguchi type coupling. The case of Kuramoto coupling has been considered in Ref.~\cite{Matteuzzi2025}, with emphasis on large ensembles. Here, we focus on a small number of coupled oscillators.

In this section, we consider the simplest possible setup of two coupled oscillators. The remarkable feature is that the reliability properties can be determined analytically. 

\subsection{Basic equations}
We restrict our attention to the simplest case of pure main harmonic coupling. Then, the general equations for coupled phase oscillators $\tilde x,\tilde y$ read
\begin{equation}
\begin{aligned}
\frac{d}{dt}\tilde x&=\tilde\w_x+\tilde a_x \sin(\tilde y-\tilde x-\tilde\alpha_x)\;,\\
\frac{d}{dt}\tilde y&=\tilde\w_y+\tilde a_y \sin(\tilde x-\tilde y-\tilde\alpha_y)\;.
\end{aligned}
\label{eq:01}
\end{equation}
There are six parameters: two natural frequencies $\tilde\w_x,\tilde\w_y$, two coupling strengths $\tilde a_{x},\tilde a_{y}$, and two phase shifts $\tilde\alpha_x,\tilde\alpha_y$. With a transformation $\tilde x=(\tilde\w_x+\tilde\w_y)/2\,t+(\tilde\alpha_x-\tilde\alpha_y)/2+x$,
$\tilde y=(\tilde\w_x+\tilde\w_y)/2\,t+y$, one can make the natural frequencies equal in modulus $\pm (\tilde\w_x-\tilde\w_y)/2$, and the phase shifts equal to $\alpha=(\tilde\alpha_x+\tilde\alpha_y)/2$. Assuming for definiteness that $\tilde\w_x>\tilde\w_y$, we rescale time by the factor $(\tilde\w_x-\tilde\w_y)/2$ and obtain a system containing three parameters only:
\begin{equation}
\begin{aligned}
\dot x&=1+\e\cos\beta \sin(y-x-\alpha)\;,\\
\dot y&=-1+ \e\sin\beta \sin( x-y-\alpha)\;.
\end{aligned}
\label{eq:m}
\end{equation}
Here instead of coupling constants $\tilde a,\tilde b$, we introduced the overall coupling strength $\e$ and the coupling asymmetry parameter $\beta$ according to $2\tilde a_x/(\tilde\w_x-\tilde\w_y)=\e\cos\beta$,
$2\tilde a_y/(\tilde\w_x-\tilde\w_y)=\e\sin\beta$. It is convenient to assume $\e>0, \beta\in (-\pi,\pi),\alpha\in(-\pi,\pi)$. The cases $\beta=0,\pi$ (where $\tilde a_y=0$) or $\beta=\pm \pi/2$ (where $\tilde a_x=0$) correspond to unidirectional coupling, the case $\beta=\pi/4$ corresponds to symmetric coupling (where $\tilde a_x=\tilde a_y)$.

\subsection{Replicas and transversal Lyapunov exponents}
Together with $x,y$ we consider replicas  $X$ of $x$ and $Y$ of $y$, which are driven by the prototypes:
\begin{equation}
\begin{aligned}
\dot X&=1+\e\cos\beta \sin(y-X-\alpha)\;,\\
\dot Y&=-1+ \e\sin\beta \sin( x-Y-\alpha)\;.
\end{aligned}
\label{eq:r}
\end{equation}
For small deviations $\delta_x=X-x$, $\delta_y=Y-y$ we have linear equations
\begin{equation}
\begin{aligned}
\dot \delta_x&=-\delta_x[\e\cos\beta \cos(y-x-\alpha)]\;,\\
\dot \delta_y&=-\delta_y[ \e\sin\beta \cos( x-y-\alpha)]\;.
\end{aligned}
\label{eq:lin}
\end{equation}
This defines two LTLEs determining the reliability of phases $x,y$:
\begin{equation}
\begin{aligned}
\lambda_x&=-\langle\e\cos\beta \cos(y-x-\alpha)\rangle\;,\\
\lambda_y&=-\langle \e\sin\beta \cos( x-y-\alpha)\rangle\;.
\end{aligned}
\label{eq:trle}
\end{equation}

\subsection{Dynamics and analytic calculation of LTLEs}
Equations~\eqref{eq:m} can be solved by introducing the phase difference  $z=x-y$, which obeys a first-order equation
\begin{equation}
\dot z=2-\e A\sin(z+\phi) \;,   
\label{eq:z}
\end{equation}
where $A\cos\phi=\cos\alpha(\cos\beta+\sin\beta)$,
$A\sin\phi=\sin\alpha(\cos\beta-\sin\beta)$,
$A^2=1+\sin 2\beta\cos 2\alpha$. The dynamics of $z$ is either periodic (if $\e A<2$) or a fixed point $\bar{z}=\arcsin\frac{2}{\e A}-\phi$
(if $\e A\geq 2$). Correspondingly, the oscillators are synchronized for $\e A\geq 2$ or are in an asynchronous, quasiperiodic regime for  $\e A<2$.

Thus, the distribution density of $z$, required for calculation of the LTLEs \eqref{eq:trle}, is
\begin{equation}
\rho(z)=\begin{cases}
\delta(z-\bar{z}) & \e A\geq 2\;,\\
\frac{\sqrt{4-\e^2 A^2}}{2\pi}\frac{1}{2-\e A\sin(z+\phi)}& \e A<2\;.
\end{cases}
\label{eq:dens}
\end{equation}
Substitution of Eq.~\eqref{eq:dens} in Eq.~\eqref{eq:trle} yields LTLEs:

\textbf{Synchronous case $\e A\geq 2$}:
\begin{equation}
\begin{aligned}
\lambda_x&=
A^{-2}[\sin2\beta\sin2\alpha-\sqrt{\e^2A^2-4}\cos\beta(\cos\beta+\sin\beta\cos2\alpha)]\;,\\
\lambda_y&=
-A^{-2}[\sin2\beta\sin2\alpha+\sqrt{\e^2A^2-4}\sin\beta(\sin\beta+\cos\beta\cos2\alpha)]\;.
\end{aligned}    
\label{eq:trsyn}
\end{equation}

\textbf{Asynchronous case $\e A< 2$}:
\begin{equation}
\begin{aligned}
\lambda_x&=(2A^2)^{-1}(2-\sqrt{4-\e^2A^2})\sin2\beta\sin2\alpha\;,\\
\lambda_y&=-\lambda_x\;.
\end{aligned}    
\label{eq:trasyn}
\end{equation}
The result of Eq.~\eqref{eq:trasyn} is quite remarkable, as it shows that in the asynchronous state, either both LTLEs vanish (this happens, e.g., for a fully asymmetric coupling with $\beta=0,\;\pi/2$) or they have opposite signs, i.e., one oscillator is reliable while the other one is not. In the synchronous case, either two LTLEs have different signs, like in the case of asynchrony, or both are negative.

\subsection{Reliability properties}
\label{sec:prr}
Because the system of Eq.~\eqref{eq:m} has three parameters, a graphical representation of all possible situations is hardly possible. 
Therefore, discuss properties of reliability in particular cases.

\paragraph{Kuramoto coupling $\alpha=0$.} In this case, both LTLEs vanish in the asynchronous case at small coupling. In the synchronous regime at strong couplings, the sum of LTLEs is always negative, i.e., they cannot be both positive. The LTLEs have opposite signs if $\sin 2\beta<0$, i.e., if the coupling coefficients $\tilde{a}_x,\tilde{a}_y$ in Eq.~\eqref{eq:01} have opposite signs.

\paragraph{Symmetric coupling $\beta=\pi/4$.} In this case, LEs generally do not vanish in both the synchronous and the asynchronous states.

\begin{figure}[!htb]
    \centering
    \includegraphics[width=\columnwidth]{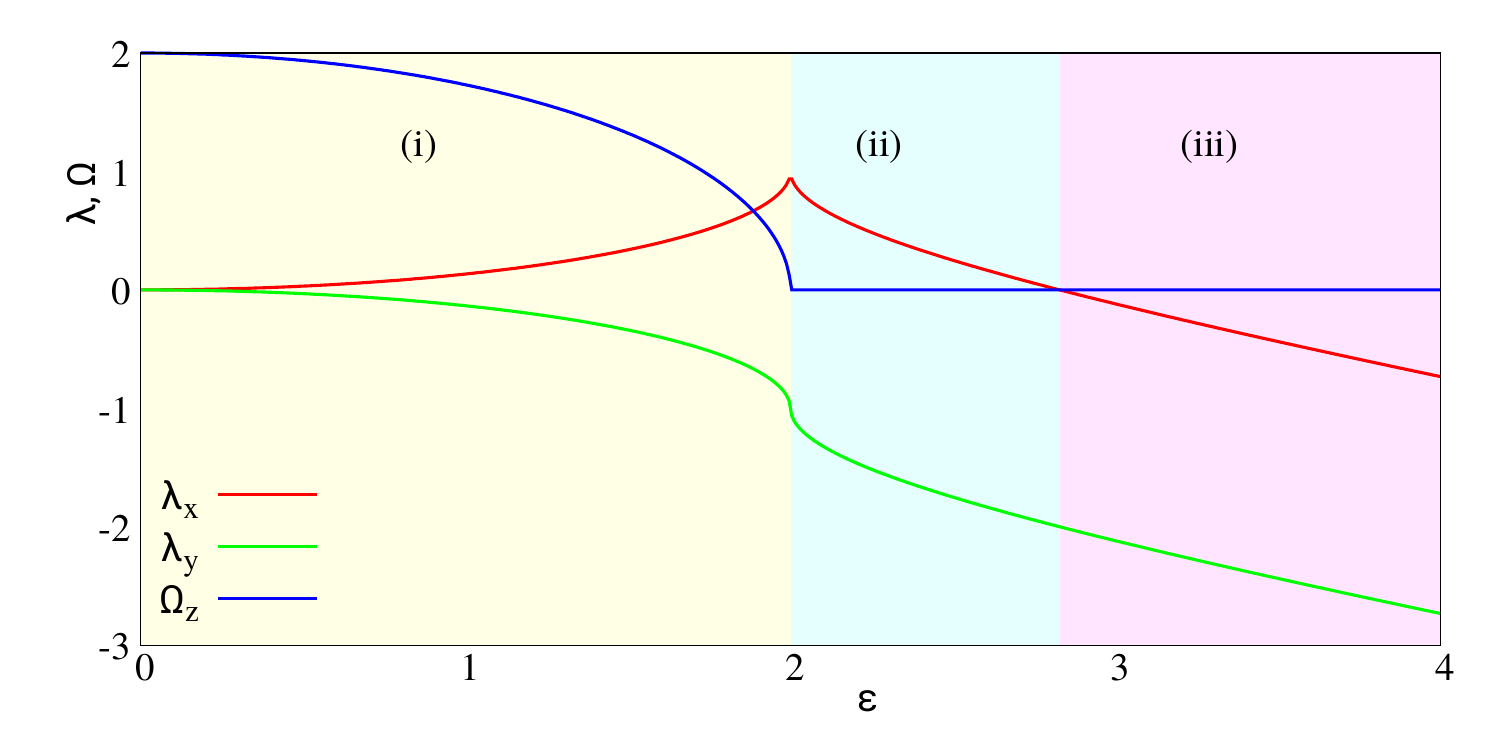}
    \caption{Transversal LEs, Eq.~\eqref{eq:sc}, as functions of the coupling strength. Additionally, we show the frequency difference $\W_z=\langle \dot z\rangle$ to indicate the synchronization transition at $\e=2$. Colored  regions indicate different states according to synchrony and reliability properties, as described in the text.}
    \label{fig:les}
\end{figure}

Let us consider, as a representative example, the case $\alpha=\beta=\pi/4$. Then $A=1$ and the expressions for the LEs are simple:
\begin{equation}
\lambda_x=\begin{cases}\frac{(2-\sqrt{4-\e^2})}{2} & \e<2\;, \\
1-\frac{\sqrt{\e^2-4}}{2} & \e\geq 2\;,
\end{cases}\qquad
\lambda_y=\begin{cases}-\frac{(2-\sqrt{4-\e^2})}{2} & \e<2\;, \\
-1-\frac{\sqrt{\e^2-4}}{2} & \e\geq 2\;.
\end{cases}
\label{eq:sc}
\end{equation}
We illustrate these LEs in Fig.~\ref{fig:les}. One can see that $\lambda_y$ is always negative, while $\lambda_x$ is positive for small couplings and becomes negative for $\e>2\sqrt{2}$. This means that the oscillator $x$ is anti-reliable in the whole asynchronous domain, and in the synchronous region for small coupling strengths.

\begin{figure}[!htb]
    \centering
    \includegraphics[width=\columnwidth]{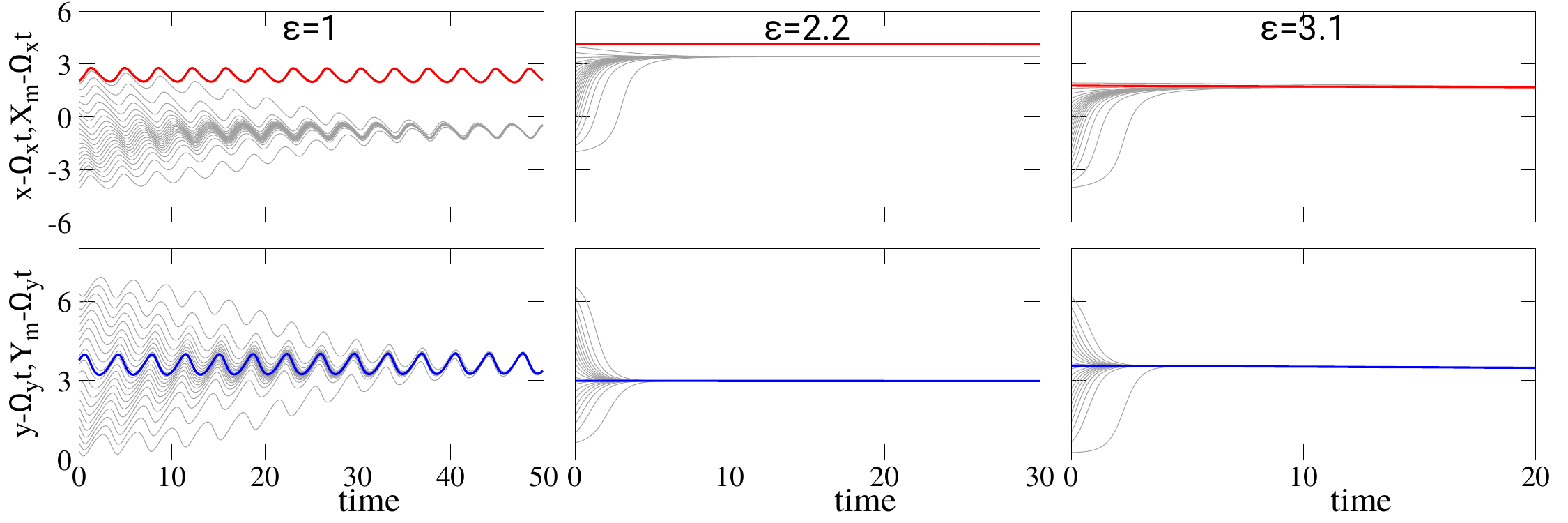}
    \caption{Prototypes (red and blue) and replicas (grey) for three values of coupling strength $\e$. \new{To enhance visibility, linear-in-time terms $\Omega_{x}t,\Omega_y t$  are subtracted from all phases on each panel, where $\Omega_{x}=\langle \dot x\rangle,\;\Omega_{y}=\langle \dot y\rangle$ are mean frequencies.} }
    \label{fig:2field}
\end{figure}

We illustrate different regimes in Fig.~\ref{fig:2field}. In these figures, we show trajectories of prototypes (colored curves) and 20 replicas (grey curves), so that one can easily see which solution the replicas converge to. In all cases, the unit $y$ is reliable. For $\e=1$ (regime (i) of Fig.~\ref{fig:les}), the dynamics is quasiperiodic and asynchronous; the unit $x$ is antireliable, and the replicas $X_m$ do not converge to the prototype $x$, while the unit $y$ is reliable \new{and the replicas $Y_m$ converge to the prototype $y$.} The same reliability properties hold for $\e=2.2$ (regime (ii) of Fig.~\ref{fig:les}), but now the dynamics is synchronous periodic. For $\e=3.1$ (regime (iii) of Fig.~\ref{fig:les}), the dynamics is synchronous, and both units are reliable.

\subsection{A simple example}

Here we present the simplest example with nontrivial reliability properties. We take two oscillators $x,y$ with equal frequencies, so that in the co-rotating reference frame the frequencies can be set to zero. Suppose that oscillator $x$ attempts to follow $y$, but $y$ attempts to be in anti-phase with $x$. In terms of Refs.~\cite{Hong-Strogatz-11,hong2011conformists}, oscillator $x$ is a ``conformist'' and oscillators $y$ is a ``contrarian''. For definiteness, suppose that the attraction of $x$ to $y$ is stronger than the repulsion of $y$ from $x$. Then, the equations for the oscillators, their replicas, and the phase difference $z=x-y$ read:
\begin{align}
\dot x&=2\sin(y-x)\label{eq:cc1}\;,\\
\dot y&=-\sin(x-y)\label{eq:cc2}\;,\\
\dot X&=2\sin(y-X)\label{eq:cc3}\;,\\
\dot Y&=-\sin(x-Y)\label{eq:cc4}\;,\\
\dot z&=-\sin z\label{eq:cc5}\;.
\end{align}
From Eq.~\eqref{eq:cc5} it follows that conformism wins and two oscillators relax on the stable fixed point $z=0$, i.e., they synchronize $(x=y)$. From Eqs.~\eqref{eq:cc3} and \eqref{eq:cc4} it follows that the replica of the conformist also synchronizes ($X=x=y$) and is reliable, while the replica of the contrarian is in anti-phase with its prototype ($Y=y+\pi$) and thus anti-reliable. In the diametrical case of stronger repulsion between $x$ and $y$, the situation will be opposite: the conformist will be anti-reliable, and the contrarian will be reliable. 

\section{Reliability in Kuramoto-Sakaguchi ensembles}
\label{sec:kse}
For three or more coupled oscillators, even if the coupling functions contain the basic harmonics only, there are many different possibilities due to different choices of the natural frequencies, the coupling strengths, and the phase shifts. Therefore, below we restrict our attention to the case of global coupling, in which all coupling strengths and phase shifts are equal.  
The oscillators differ only by their frequencies, and we will assume a uniform distribution of frequencies: for $N$ units, the set of frequencies is $-1,-1+2/(N-1),\ldots,1$. The example of Section~\ref{sec:prr} belongs to this class. The Kuramoto ensemble case with $\alpha=0$ has been explored in Ref.~\cite{Matteuzzi2025}; here we focus on the Kuramoto-Sakaguchi (KS) ensemble with $\alpha=\pi/4$, for several representative values of $N$.

\subsection{Transversal exponents}
Here we report on reliability properties of the KS ensemble with phase shift $\alpha=\pi/4$:
\begin{equation}
\dot x_j=\w_j+\frac{\e}{N}\sum_{l\neq j}\sin(x_l-x_j-\pi/4)\;,\quad j=1,\ldots,N\;.
    \label{eq:eks}
\end{equation}
According to the general expression of Eq.~\eqref{eq:bas3}, the dynamics of the transversal perturbation $u_k$ for each unit is one-dimensional. Thus, the LTLE (in fact, here there is only one transversal exponent) reduces to the time averaging of the corresponding factor:
\begin{equation}
\lambda_k=-\frac{\e}{N} \sum_{l\neq k} \langle \cos(x_l-x_k-\pi/4)\rangle\;.
    \label{eq:kstr}
\end{equation}
The system of two oscillators, analytically solved in Sec.~\ref{sec:prr}, belongs to this family (with slightly different normalization of the coupling constant).

\begin{figure}
    \centering
    \includegraphics[width=\linewidth]{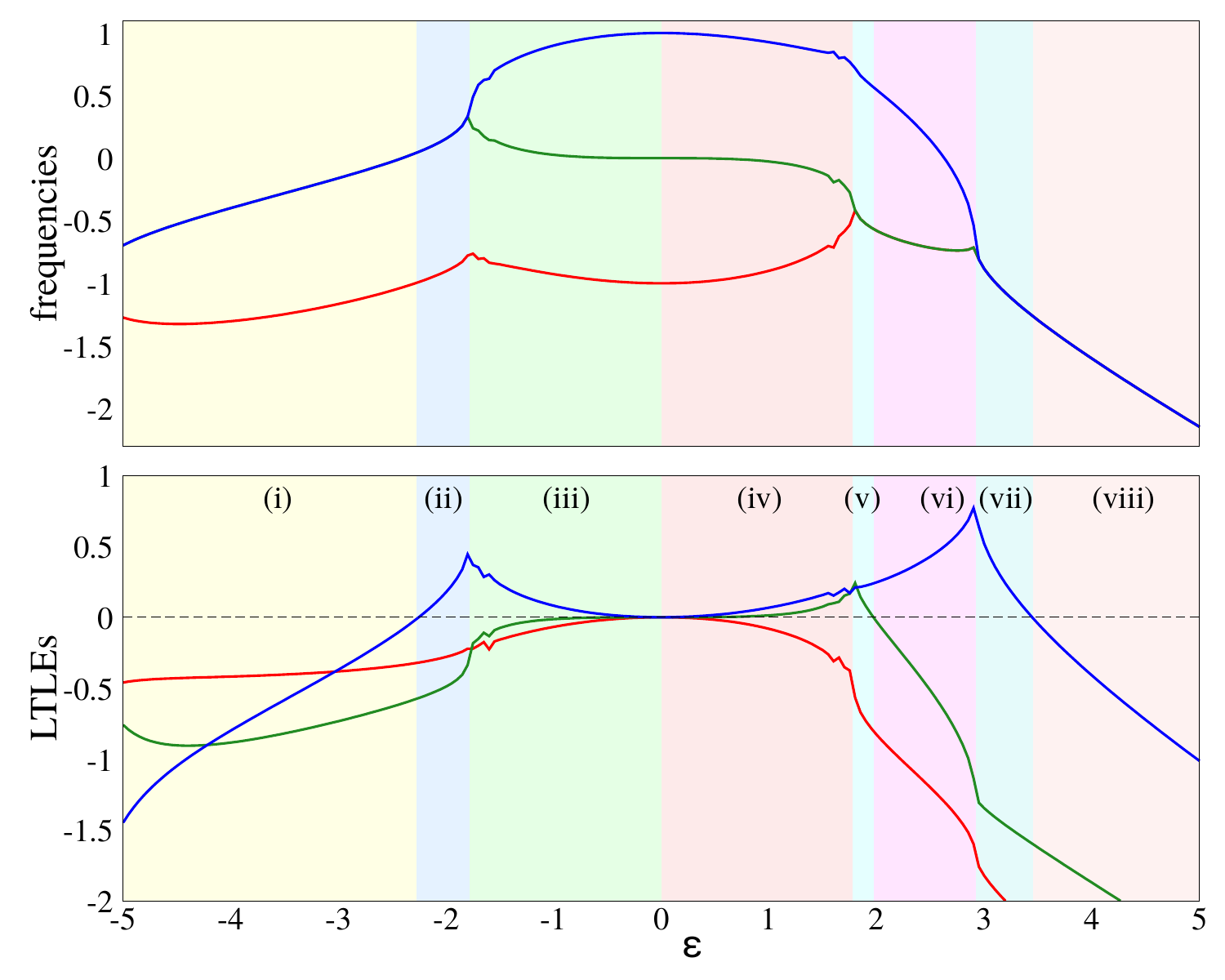}
    \caption{Observed frequencies (upper panel) and LTLEs (bottom panel) for $N=3$ oscillators with Kuramoto-Sakaguchi coupling (Eq.~\eqref{eq:eks}). Red, green, and blue lines depict oscillators with natural frequencies $-1$, $0$, $1$, respectively. Colored  regions indicate different states according to synchrony and reliability properties, as described in detail in the text.}
    \label{fig:ks03}
\end{figure}

We illustrate the case $N=3$ in Fig.~\ref{fig:ks03}. Here, there are eight different regimes, which can be classified according to properties of synchrony (according to frequencies in the upper panel) and to the properties of reliability (via LTLEs in the bottom panel). Here for small $|\e|$ all units have different frequencies (regimes (iii),(iv)), for larger $|\e|$ a two-frequency regime is observed (regimes (i),(ii),(v),(vi)), and for large positive $\e$ all oscillators are synchronized and possess the same frequency (regimes (vii),(viii)). All these states are regular, with the vanishing largest Lyapunov exponent. The oscillator with the lowest natural frequency is reliable for all coupling strengths, the oscillator with the middle natural frequency is anti-reliable for small positive values of $\e$ (regimes (iv),(v)), and the oscillator with the largest natural frequency is anti-reliable both for positive and negative couplings (regimes (ii),(iii),(iv),(v),(vi),(vii)), while the domain of anti-reliability is larger for positive $\e$. This oscillator is reliable in domains (i),(viii). Similar to the case of two oscillators, Fig.~\ref{fig:les}, there is a domain (vii) of coupling strengths around $\e\approx 3.2$, where in the regime of full synchrony, the unit with the largest natural frequency is anti-reliable. 

\begin{figure}
    \centering
    \includegraphics[width=0.49\linewidth]{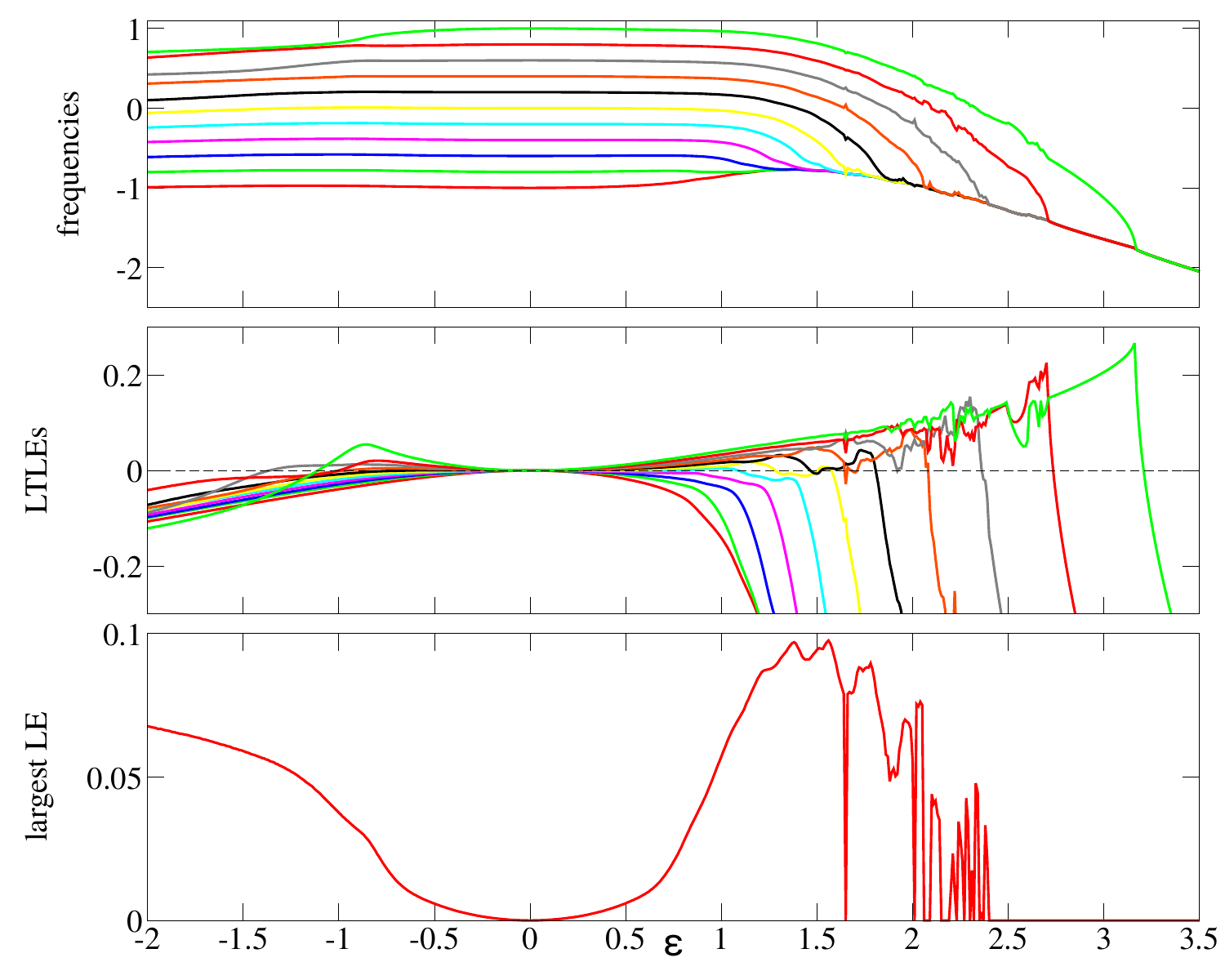}\hfill
    \includegraphics[width=0.49\linewidth]{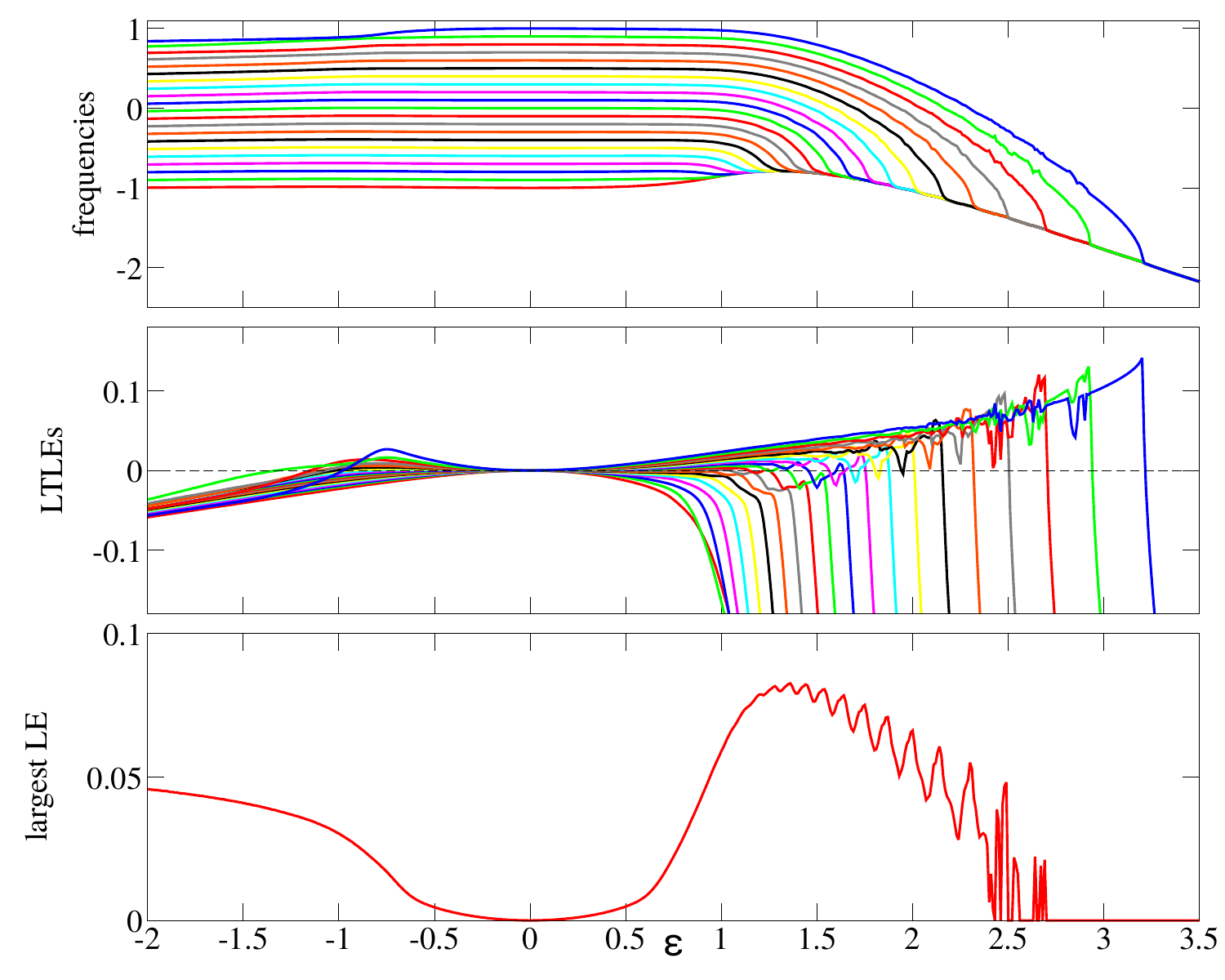}
    \caption{Reliability properties of the KS ensembles, Eq.~\eqref{eq:eks}, with $N=11$ (left) and $N=21$ (right). Upper panels: observed frequencies vs $\e$, showing a synchronization transition at positive $\e$. Middle panels: LTLEs revealing reliable (negative values) and anti-reliable (positive values) oscillators. Bottom panels: the largest Lyapunov exponent of the dynamics.}
    \label{fig:ks1121}
\end{figure}

The cases with larger numbers of units in an ensemble are illustrated in Fig.~\ref{fig:ks1121}. The first observation is that in a large domain of coupling strengths, the dynamics is chaotic~\cite{Popovych-Maistrenko-Tass-05,maistrenko2005chaotic,Carlu-Ginelli-Politi-18,brister2020three}, as the presented largest Lyapunov exponent of Eq.~\eqref{eq:eks} shows (bottom panels). Only for large positive $\e$, where there is a large cluster of synchronized units with a few having maximal natural frequency that still deviate in the observed frequencies, the dynamics becomes regular. The second observation is that anti-reliability is characteristic for oscillators with larger natural frequencies. So, for $N=11$, the units $1\leq k\leq 4$ are always reliable; units $k=5,6$ are anti-reliable in some range of positive couplings; and units $7\leq k\leq 11$ are anti-reliable for positive and negative couplings. The third observation is that reliability is not directly related to chaos; anti-reliable and reliable units are observed in both chaotic and regular states.  

\subsection{Illustration of reliable and anti-reliable dynamics}

\begin{figure}
    \centering
    \includegraphics[width=0.8\linewidth]{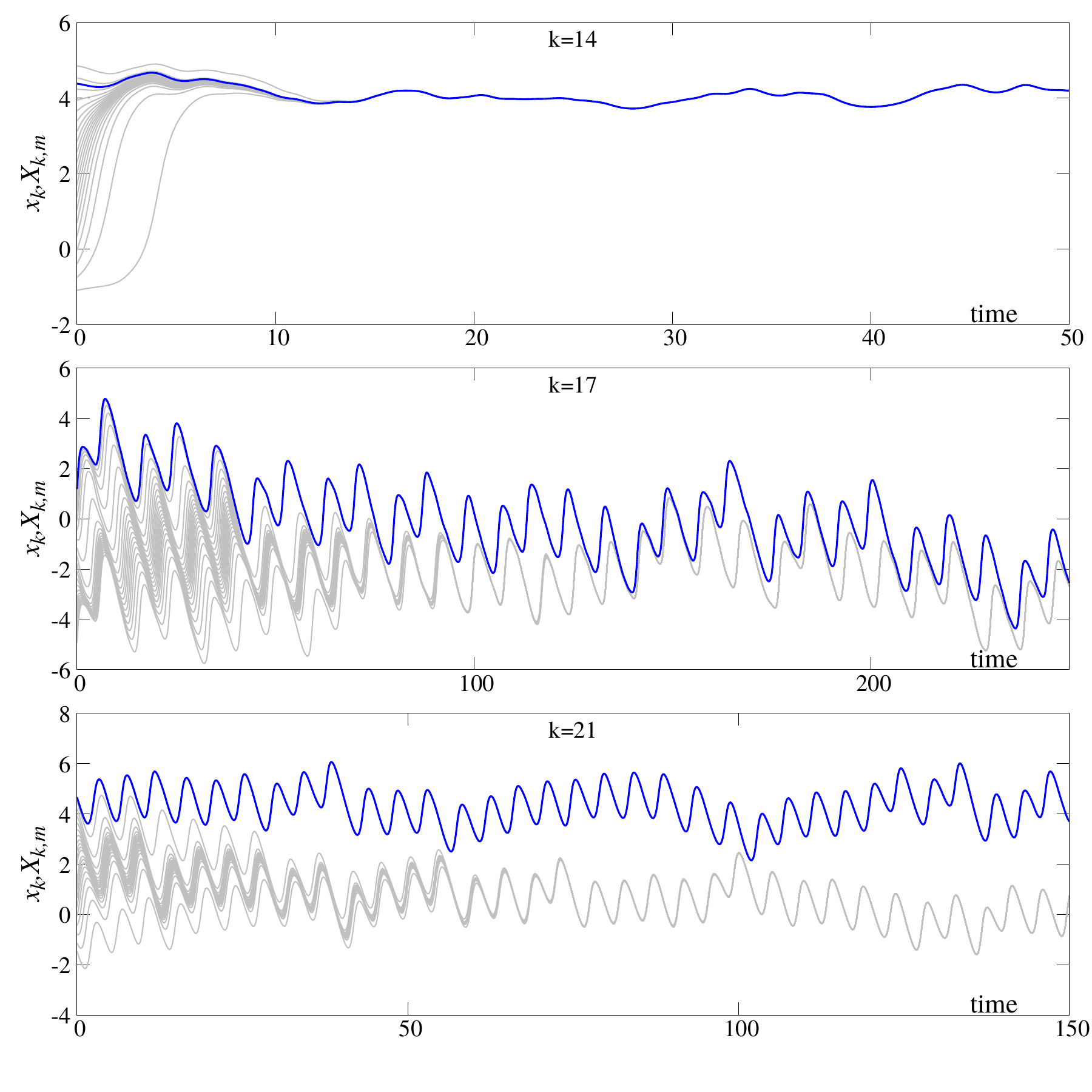}
    \caption{Evolution of the prototypes $x_k$ (blue curves) and of 20 replicas $X_{k,m}$ (grey curves) in the Kuramoto-Sakaguchi model, Eq.~\eqref{eq:eks}, for $\e=2$ and $N=21$. The unit $k=14$ is reliable, while units $k=17,21$ are anti-reliable. For clarity of presentation, from all graphs, the average phase shift $\Omega_k t$ is subtracted.}
    \label{fig:ar}
\end{figure}

In Fig.~\ref{fig:ar} we illustrate reliable and anti-reliable dynamics in the KS model of Eq.~\eqref{eq:eks} for $\e=2$ and $N=21$. We show 20 replicas of 3 units: (i) $k=14$, this unit belongs to the cluster of synchronized oscillators and is reliable \new{(all replicas $X_{k,m}$ converge to the prototype $x_k$)}; (ii) $k=17$, this unit is anti-reliable with relatively small positive transversal LE \new{(all replicas $X_{k,m}$ converge to a trajectory which is different from the prototype $x_k$ but remains relatively close to it)}; (iii) $k=21$, this unit is strongly anti-reliable \new{(all replicas $X_{k,m}$ converge to a trajectory which is different from the prototype $x_k$ and lies far away from it.)}.

\subsection{Properties of a replica-attractor and a replica-repeller in the KS model.}

We note that in each panel of Fig.~\ref{fig:ar}, all the units are driven by the same field, only their initial conditions at $t=0$ differ. The state of the prototype (blue curves) is fixed by the preceding dynamics of the full system of Eq.~\eqref{eq:eks}, while initial states of replicas can be chosen arbitrarily. In all cases, the replicas converge to a single trajectory, which we dub \textit{replica-attractor}. The trajectory of the prototype is also replica-attractor if the transversal LE is negative (case $k=14$ in Fig.~\ref{fig:ar}), but it is \textit{replica-repeller} if the transversal LE is positive (cases $k=17,k=21$ in Fig.~\ref{fig:ar}). We stress here that the prototype always belongs to an attractor in the full system, Eq.~\eqref{eq:eks}, thus we use different terms to emphasize that Fig.~\ref{fig:ar} illustrates attraction and repulsion in the replica space. 

Replica-attractor and replica-repeller in the KS system can be further characterized by virtue of the Watanabe-Strogatz (WS) theory~\cite{Watanabe-Strogatz-93,Watanabe-Strogatz-94}. In particular, we use a complex formulation of the WS theory according to \cite{Marvel-Mirollo-Strogatz-09,Pikovsky-Rosenblum-15}, and Hamiltonian formulation according to \cite{Braun-etal-12}.

Here we discuss in detail the dynamics of replicas of a particular unit $k$ in the ensemble, Eq.~\eqref{eq:eks}. For brevity of notations, we omit index $k$ everywhere, and denote the set of replicas as $\vp_j(t)$, where different $j$ correspond to different initial conditions with which these replicas start at $t=0$.

To apply WS theory, it is convenient to write equations for $\vp_j$ in complex form
\begin{equation}
\dot\vp_m=\w+\frac{\e}{N}\sum_{l\neq k}\sin(x_l-\vp_m-\alpha)=\w+\text{Im}(H(t)e^{-i\vp_m}),\quad H(t)=\frac{\e}{N}\sum_{l\neq k} e^{i(x_l-\alpha)}\;.
\label{eq:ws}
\end{equation}
According to the WS theory, one introduces new variables $\psi_j$ (WS phases) and complex WS amplitude $z$ according to
\begin{equation}
e^{i\vp_m}=\frac{z+e^{i\psi_m}}{1+z^*e^{i\psi_m}}\;.
\label{eq:ws2}
\end{equation}
Then, the WS equations for $z,\psi_j$ are
\begin{align}
\dot z&=i\w z+\frac{1}{2}(H-H^*z^2)\;,
\label{eq:ws3}\\
\dot\psi_m&=\w+\text{Im}(z^*H)\;.
\label{eq:ws4}
\end{align}
One can see that the relation of Eq.~\eqref{eq:ws4} for the WS phases are all the same, which means that there is in fact one nontrivial equation, and the differences of the WS phases are constants of motion (partial integrability).

The transformation of Eq.~\eqref{eq:ws2} is a M\"obius transformation, and it is underdetermined. One can impose an additional condition at time $t=0$. It appears convenient to set $z(0)=0$, so that $\psi_m(0)=\vp_m(0)$. If one introduces a variable $\Psi$ which obeys Eq.~\eqref{eq:ws4} with initial condition $\Psi(0)=0$, i.e.,
\begin{equation}
\dot\Psi=\w+\text{Im}(z^*H)\;,
\label{eq:ws5}
\end{equation}
then $\psi_m(t)=\vp_m(0)+\Psi(t)$. This gives a solution of the system of Eq.~\eqref{eq:ws} through the dynamics of $z,\Psi$:
\begin{equation}
e^{i\vp_m(t)}=\frac{z(t)+e^{i(\vp_m(0)+\Psi(t))}}{1+z^*(t)e^{i(\vp_m(0)+\Psi(t))}}\;.
\label{eq:ws6}
\end{equation}

It is more convenient to work with real variables $J,\beta,\gamma$, which we define as
\begin{gather*}
H(t)=h(t)\exp[i\Phi(t)],\quad z=\rho\exp[i(\Phi+\beta)],\quad \Psi=\Phi+\beta+\gamma,\quad
J=\frac{\rho^2}{2(1-\rho^2)},\quad \rho=\sqrt{\frac{2J}{1+2J}}\;,
\end{gather*}
and rewrite the system of Eqs.~\eqref{eq:ws3} and\eqref{eq:ws5} together with Eq.~\eqref{eq:ws6} as a two-dimensional Hamiltonian system (where $J$ is action and $\beta$ is angle)
\begin{equation}
\begin{aligned}
\dot J&=\frac{\sqrt{2J(1+2J)}}{2}h(t)\cos(\beta)=-\frac{\partial \mathcal{H}(J,\beta)}{\partial \beta}\;,\\
\dot\beta&=\w-\dot\Phi(t)-\frac{1+4J}{2\sqrt{2J(1+2J)}}h(t)\sin(\beta)=\frac{\partial \mathcal{H}(J,\beta)}{\partial J}\;,\\
\mathcal{H}(J,\beta)&=J[\w-\dot\Phi(t)]-h(t)\frac{\sqrt{2J(1+2J)}}{2}\sin\beta\;,\\\
\dot\gamma&=\frac{1}{2\sqrt{2J(1+2J)}} h(t)\sin(\beta)\;,\\
\exp[i\vp_m(t)]&=\exp[i(\beta+\Phi(t))]\frac{\sqrt{\frac{2J}{1+2J}}+\exp[i(\vp_m(0)+\gamma)]}{1+\sqrt{\frac{2J}{1+2J}} \exp[i(\vp_m(0)+\gamma)]}\;.
\end{aligned}
\label{eq:ws10}
\end{equation}

Let us focus on the dynamics at large times, where $J\gg 1$  (what corresponds to $\rho\approx 1$). In this limit the system of Eq.~\eqref{eq:ws10} reduces to
\begin{align}
\dot J&=Jh(t)\cos\beta=-\frac{\partial \tilde{\mathcal{H}}(J,\beta)}{\partial \beta}\label{eq:ws11-1}\;,\\
\dot\beta&=\w-\dot\Phi(t)-h(t)\sin\beta=\frac{\partial \tilde{\mathcal{H}}(J,\beta)}{\partial J}\label{eq:ws11-2}\;,\\
\tilde{\mathcal{H}}(J,\beta)&=J(\w-\dot\Phi(t)-h(t)\sin\beta)\label{eq:ws11-3}\;,\\
\dot\gamma&=\frac{1}{4J} h(t)\sin(\beta)\label{eq:ws11-4}\;,\\
\tan\frac{\vp_m(t)-\beta(t)-\Phi(t)}{2}&=(8J)^{-1}\tan\frac{\vp_m(0)+\gamma(t)}{2}\label{eq:ws11-5}\;,
\end{align}
where we used a formula $\exp[ia]=
(1+i\tan(a/2))(1-i\tan(a/2))^{-1}$ to write the M\"obius transformation $\vp_m(0)\to\vp_m(t)$ in the real form.

Let us discuss properties of this solution, and implications for replica-attractor and replica-repeller.
\begin{enumerate}
\item We assume that $J\to\infty$ (or $\rho\to 1$) as $t\to\infty$, this is the condition for the existence of an attractor and a repeller. The growth rate of $J$ follows from Eq.~\eqref{eq:ws11-1}: $J\sim \exp[\mu t]$, where $\mu= \langle h(t)\cos\beta(t)\rangle$.
\item At large times where $J\gg 1$, according to Eq.~\eqref{eq:ws11-5}, almost all initial phases $\vp_j(0)$ converge to the attractor $\vp(t)=\beta(t)+\Phi(t)$. 
\item However, convergence to the attractor is ensured only outside a small vicinity of the point $\vp_m(0)+\gamma(t)=\pi$. One can see that as $t\to \infty$, according to Eq.~\eqref{eq:ws11-4}, $\gamma(t)\to \Gamma$ because the r.h.s. of Eq.~\eqref{eq:ws11-4} tends exponentially to zero. Thus, the initial condition $\vp_m(0)=\pi-\Gamma$ does not converge to a replica-attractor, and therefore it produces a replica-repeller. Note that the exact initial position of the repeller can be obtained only after the whole trajectory for all $t>0$ is known.
\item Stability of the attractor can be calculated by taking the derivative of the transformation of Eq.~\eqref{eq:ws11-5}:
\begin{equation}
\begin{gathered}
\frac{d\vp_m(t)}{d\vp_m(0)}=(8J)^{-1}\frac{\cos^2\frac{\vp_m(t)-\beta(t)-\Phi(t)}{2}}{\cos^2 \frac{\vp_m(0)+\gamma(t)}{2}}=
\frac{1+\tan^2\frac{\vp_m(0)+\gamma(t)}{2} }{8J+(8J)^{-1} \tan^2\frac{\vp_m(0)+\gamma(t)}{2} }
\end{gathered}
\label{eq:ws16}
\end{equation}
One can see that for all $\vp_m(0)$ that are not close to $\pi-\gamma(t)$, the derivative $\frac{d\vp_m(t)}{d\vp_m(0)}\sim J^{-1}$.
Defining the average stability of the trajectory (the negative Lyapunov exponent of the attractor) as $\Lambda_{\text{attr}}=\lim_{t\to\infty} \frac{1}{t}\frac{d\vp_j(t)}{d\vp_m(0)}$, and taking into account that $J\sim \exp[\mu t]$, we get $\Lambda_{\text{attr}}=-\mu=-\langle h(t)\cos\beta(t)\rangle$.
\item The instability of the repeller can be calculated by noting that the M\"obius map, Eq.~\eqref{eq:ws11-5}, can be inverted. We can simply invert the derivative of Eq.~\eqref{eq:ws16}:
\begin{equation}
\begin{gathered}
\frac{d\vp_m(0)}{d\vp_m(t)}=(8J)\frac{\cos^2 \frac{\vp_m(0)+\gamma(t)}{2}}{\cos^2\frac{\vp_m(t)-\beta(t)-\Phi(t)}{2}}=
(8J)^{-1}\frac{1+\tan^2\frac{\vp_m(t)-\beta(t)-\Phi(t)}{2}}{(8J)^{-2}+\tan^2\frac{\vp_m(t)-\beta(t)-\Phi(t)}{2}}
\end{gathered}
\label{eq:ws17}
\end{equation}
One can see that for all $\vp_j(t)$ that are not close to $\beta(t)+\Phi(t)$ (i.e. are not close to the replica-attractor at time $t$), the derivative of the backward evolution $\frac{d\vp_m(0)}{d\vp_m(t)}\sim J^{-1}$. Because the evolution backward in time is concentrated at the replica-repeller, this shows that the Lyapunov exponent on the replica-repeller is $\Lambda_{\text{rep}}=\mu=\langle h(t)\cos\beta(t)\rangle$. Comparing with Eqs.~\eqref{eq:ws11-1},\eqref{eq:ws11-2}, one can see that $\lambda$ and $\Lambda_{\text{attr}}=-\Lambda_{\text{rep}}$ are two Lyapunov exponents of a Hamiltonian system, and their symmetry corresponds to the general symmetry of LEs in Hamiltonian dynamics.
\end{enumerate}
Let us now discuss how the properties of the attractor and the repeller resulting from the WS theory are related to the originally introduced properties of the prototype and replicas. For this we rewrite the dynamics of the prototype $x_k$ from Eq.~\eqref{eq:eks} and its replica from Eq.~\eqref{eq:ws}, again omitting index $k$ for simplicity
\begin{align}
\dot x&=\w+h(t)\sin(\Phi(t)-x) \label{eq:pr1}\;,\\
\dot\vp&=\w+h(t)\sin(\Phi(t)-\vp)\label{eq:pr2}\;,\\
he^{i\Phi}&=\frac{\e}{N}\sum_{l\neq k} e^{i(x_l-\alpha)}\;.
\end{align}
Replica $\vp$, initial condition of which is chosen ``randomly'', almost always converges to an attractor, on which $\vp(t)-\Phi(t)=\beta(t)$. Thus, stability of the replica is always given by $\Lambda_{\text{attr}}=\langle h(t)\cos(\vp(t)-\Phi(t))\rangle$.

\begin{figure}
    \centering
    \includegraphics[width=0.5\linewidth]{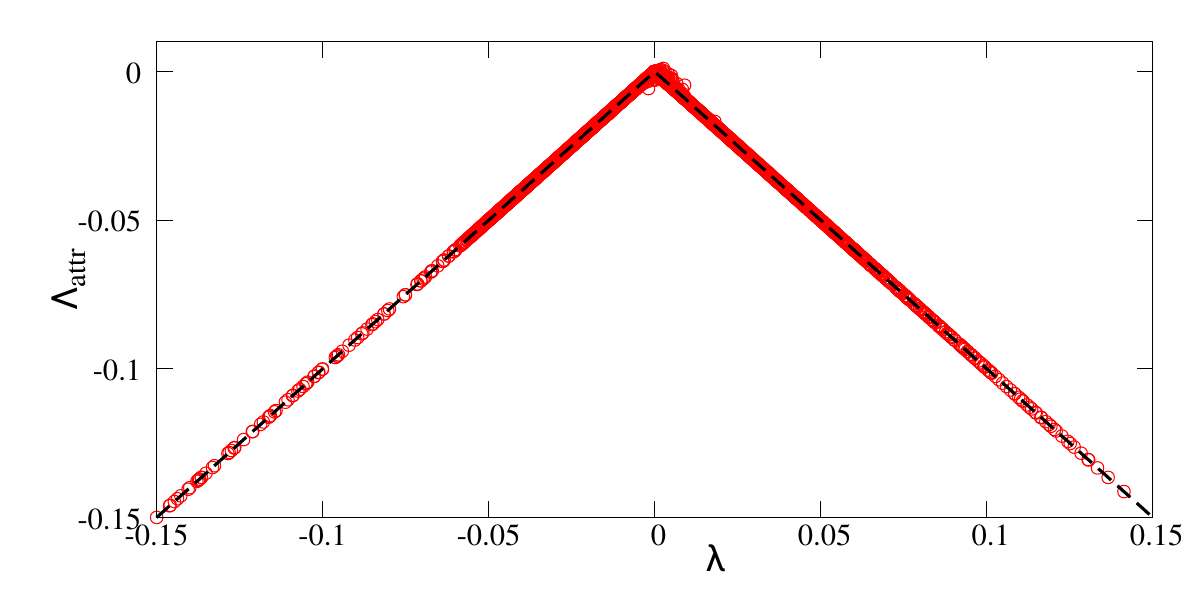}
    \caption{Test of the relation of Eq.~\eqref{eq:pr3} in a Kuramoto-Sakaguchi ensemble. Circles: calculated values of $\lambda$ and $\Lambda_{\text{attr}}$; black dashed line: the relation of Eq.~\eqref{eq:pr3}.}
    \label{fig:sym}
\end{figure}

Prototype $x$ can lie on the repeller or on the attractor (in this case $x(t)\to\vp(t)$ as $t\to\infty$). In the former case, it is unstable, and its transversal LE is $\lambda=-\langle h(t)\cos(x(t)-\Phi(t))\rangle=\Lambda_{\text{rep}}=\mu>0$. In the latter case, it is stable, and the transversal LE is $\lambda=-\langle h(t)\cos(x(t)-\Phi(t))\rangle=-\langle h(t)\cos(\vp(t)-\Phi(t))\rangle=\Lambda_{\text{attr}}=-\mu>0$. Summarizing, we obtain the following symmetry relation between the stability of the replica $\Lambda_{\text{attr}}$ and the LTLE of the prototype $\lambda$:
\begin{equation}
\Lambda_{\text{attr}}=-|\lambda|\;.
    \label{eq:pr3}
\end{equation}
We illustrate this relation in Fig.~\ref{fig:sym}. Here we show values $\lambda$ and $\Lambda_{\text{attr}}$ for all units in an ensemble of $N=21$ Kuramoto-Sakaguchi oscillators, Eq.~\eqref{eq:eks}, for all values of $\e$ in the interval presented in Fig.~\ref{fig:ks1121} (only cases with 
$|\lambda|<0.15$ are chosen for better visibility).

\section{Conclusions}
\label{sec:concl}

In summary, we have explored properties of internal reliability in a network of phase oscillators coupled via Kuramoto-Sakaguchi-type coupling. Reliability characterizes the ``stability'' of the dynamics with respect to replicas of particular units; reliability means that all the replicas converge toward the prototype, while in an anti-reliable case, replicas converge to a different state. The presence of reliable and anti-reliable units appears to be a generic feature in a wide range of system sizes and dynamical regimes. For two coupled oscillators, reliability properties have been calculated analytically. Here, in a general situation, in the asynchronous state, one oscillator is reliable and another one anti-reliable.  This situation can also be observed in the synchronous state, although in some regions of parameters, both oscillators are reliable. 

The case of Kuramoto-Sakaguchi phase oscillators is particular, because a set of replicas possesses one attracting and one repelling trajectory. We speak here about ``replica-attractor'' and ``replica-repeller'', not to be confused with attractors and repellers in the original system.  Therefore, one can interpret a reliable unit as one staying on the replica-attractor, while in an anti-reliable case, the prototype stays on the replica-repeller. Furthermore, there is an interesting symmetry between replica-attractor and replica-repeller following from the Watanabe-Strogatz description of the dynamics of Kuramoto-Sakaguchi oscillators: their stability exponents have the same value but opposite signs.

%
%

\section*{Acknowledgements}
FB wishes to thank the University of Perpignan Via Domitia for kind hospitality. 
SI acknowledges financial support from the Italian MUR PRIN2022 project ``Breakdown of ergodicity in classical and quantum many-body systems" (BECQuMB) Grant No.
	20222BHC9Z.






\bibliographystyle{vancouver}

\bibliography{reliability}

@article{Matteuzzi2025,
  title = {Internal reliability and antireliability in dynamical networks},
  volume = {112},
  ISSN = {2470-0053},
  DOI = {10.1103/p5xz-rfv6},
  number = {4},
  journal = {Physical Review E},
  publisher = {American Physical Society (APS)},
  author = {Matteuzzi,  Tommaso and Bagnoli,  Franco and Baia,  Michele and Iubini,  Stefano and Pikovsky,  Arkady},
  year = {2025},
  month = oct 
}

@article{Pecora1990,
  title = {Synchronization in chaotic systems},
  volume = {64},
  ISSN = {0031-9007},
  DOI = {10.1103/physrevlett.64.821},
  number = {8},
  journal = {Physical Review Letters},
  publisher = {American Physical Society (APS)},
  author = {Pecora,  Louis M. and Carroll,  Thomas L.},
  year = {1990},
  month = feb,
  pages = {821–824}
}

@ARTICLE{Mainen-Sejnowski-95
 ,AUTHOR=  {Z. F. Mainen and T. J. Sejnowski
 },TITLE=  {Reliability of Spike Timing in Neocortical Neurons
 },JOURNAL={Science
 },YEAR=   {1995
 },VOLUME= {268
 },NUMBER= {
 },PAGES=  {1503
}  }

@article{Goldobin-Pikovsky-06b,
	author = {D. S. Goldobin and A. Pikovsky},
	title = {Antireliability of noise-driven neurons},
	journal = {Phys. Rev. E},
	year = {2006},
	volume = {73},
	pages = {061906}}

@article{Pikovsky-84a,
	author = {A. S. Pikovsky},
	title = {Synchronization and stochastization of the ensemble of
autogenerators by external noise},
	journal = {Radiophys. Quantum Electron.},
	year = {1984},
	volume = {27},
	number = {5},
	pages = {390-395}}

@ARTICLE{Rulkov-Sushchik-Tsimring-Abarbanel-95
 ,AUTHOR=  {N. F. Rulkov and M. M. Suschik and 
L. S. Tsimring and H. D. I. Abarbanel
 },TITLE=  {Generalized synchronization of chaos in directionally coupled chaotic systems
 },JOURNAL={Phys.  Rev.  E
 },YEAR=   {1995
 },VOLUME= {51
 },NUMBER= {
 },PAGES=  {980
 }  }

@article{Carlu-Ginelli-Politi-18,
  title = {Origin and scaling of chaos in weakly coupled phase oscillators},
  author = {Carlu, Mallory and Ginelli, Francesco and Politi, Antonio},
  journal = {Phys. Rev. E},
  volume = {97},
  issue = {1},
  pages = {012203},
  numpages = {11},
  year = {2018},
  month = {Jan},
  publisher = {American Physical Society},
}

@article{Popovych-Maistrenko-Tass-05,
  title = {Phase chaos in coupled oscillators},
  author = {Popovych, Oleksandr V. and Maistrenko, Yuri L. and Tass, Peter A.},
  journal = {Phys. Rev. E},
  volume = {71},
  issue = {6},
  pages = {065201},
  numpages = {4},
  year = {2005},
  month = {Jun},
  publisher = {American Physical Society},
}

@article{ermentrout2008reliability,
  title={Reliability, synchrony and noise},
  author={Ermentrout, G Bard and Gal{\'a}n, Roberto F and Urban, Nathaniel N},
  journal={Trends in neurosciences},
  volume={31},
  number={8},
  pages={428--434},
  year={2008},
  publisher={Elsevier}
}

@article{lin2009reliability,
  title={Reliability of coupled oscillators},
  author={Lin, Kevin K and Shea-Brown, Eric and Young, Lai-Sang},
  journal={Journal of nonlinear science},
  volume={19},
  pages={497--545},
  year={2009},
  publisher={Springer}
}

@Article{Watanabe-Strogatz-94,
  author =       "S. Watanabe and S. H. Strogatz",
  title =        "Constants of motion for superconducting {J}osephson arrays",
  journal =      "Physica D",
  volume =       "74",
  number =       "",
  pages =        "197-253",
  year =         "1994",
}

@Article{Watanabe-Strogatz-93,
  author =       "S. Watanabe and S. H. Strogatz",
  title =        "Integrability of a globally coupled oscillator array",
  journal =      "Phys. Rev. Lett.",
  volume =       "70",
  number =       "16",
  pages =        "2391-2394",
  year =         "1993",
}

@article{Baia2025,
  title = {Synchronization of branching chain of dynamical systems},
  volume = {477},
  ISSN = {0167-2789},
  DOI = {10.1016/j.physd.2025.134664},
  journal = {Physica D: Nonlinear Phenomena},
  publisher = {Elsevier BV},
  author = {Baia,  Michele and Bagnoli,  Franco and Matteuzzi,  Tommaso and Pikovsky,  Arkady},
  year = {2025},
  month = jul,
  pages = {134664}
}

@article{abarbanel1996generalized,
  title={Generalized synchronization of chaos: The auxiliary system approach},
  author={Abarbanel, Henry DI and Rulkov, Nikolai F and Sushchik, Mikhail M},
  journal={Phys. Rev. E},
  volume={53},
  number={5},
  pages={4528},
  year={1996},
  publisher={APS}
}

@article{Letelier-23,
    author = {Letellier, Christophe and Sendiña-Nadal, Irene and Leyva, I. and Barbot, Jean-Pierre},
    title = {Generalized synchronization mediated by a flat coupling between structurally nonequivalent chaotic systems},
    journal = {Chaos},
    volume = {33},
    number = {9},
    pages = {093117},
    year = {2023},
    month = {09},
    issn = {1054-1500},
    doi = {10.1063/5.0156025}
}

@article{letellier2024taxonomy,
  title={A taxonomy for generalized synchronization between flat-coupled systems},
  author={Letellier, Christophe and Minati, Ludovico and Sendina-Nadal, Irene and Leyva, I},
  journal={arXiv preprint arXiv:2401.11561},
  year={2024}
}

@book{pikovsky2016lyapunov,
  title={Lyapunov exponents: a tool to explore complex dynamics},
  author={Pikovsky, Arkady and Politi, Antonio},
  year={2016},
  publisher={Cambridge University Press}
}

@ARTICLE{Sakaguchi-Kuramoto-86
 ,AUTHOR=  {H. Sakaguchi and Y. Kuramoto
 },TITLE=  {A soluble active rotator model showing phase transition via mutual
 entrainment
 },JOURNAL={Prog.  Theor.  Phys. 
 },YEAR=   {1986
 },VOLUME= {76
 },NUMBER= {3
 },PAGES=  {576-581
 }  }

@article{Hong-Strogatz-11,
  title = {Kuramoto Model of Coupled Oscillators with Positive and Negative Coupling Parameters: An Example of Conformist and Contrarian Oscillators},
  author = {Hong, H. and Strogatz, S. H.},
  journal = {Phys. Rev. Lett.},
  volume = {106},
  issue = {5},
  pages = {054102},
  numpages = {4},
  year = {2011},
  month = {Feb},
  doi = {10.1103/PhysRevLett.106.054102},
  publisher = {American Physical Society}
}

@article{hong2011conformists,
  title={Conformists and contrarians in a Kuramoto model with identical natural frequencies},
  author={Hong, Hyunsuk and Strogatz, Steven H},
  journal={Physical Review E—Statistical, Nonlinear, and Soft Matter Physics},
  volume={84},
  number={4},
  pages={046202},
  year={2011},
  publisher={APS}
}

@ARTICLE{Marvel-Mirollo-Strogatz-09
 ,AUTHOR=  {S. A. Marvel and R. E. Mirollo and S. H. Strogatz
 },TITLE=  {Phase oscillators with global sinusoidal coupling 
            evolve by {M}obius group action
 },JOURNAL={Chaos
 },YEAR=   {2009
 },VOLUME= {19
 },NUMBER= {
 },PAGES=  {043104.
 },NOTE=   {
}  }

@article{Pikovsky-Rosenblum-15,
   author = "Pikovsky, A. and Rosenblum, M.",
   title = "Dynamics of globally coupled oscillators: Progress and perspectives",
   journal = "Chaos",
   year = "2015",
   volume = "25",
   number = "9", 
   eid = 097616,
   pages = "097616"
}

@article{Braun-etal-12,
	author = {W. Braun and A. Pikovsky and M. A. Matias and P. Colet },
	title = {Global dynamics of oscillator populations under common noise },
	journal = {EPL},
	year = {2012},
	volume = {99},
	pages = {20006},
	issue = {},
	numpages = {}
}

@article{stanley1987dynamics,
  title={Dynamics of spreading phenomena in two-dimensional Ising models},
  author={Stanley, H Eugene and Stauffer, Dietrich and Kertesz, Janos and Herrmann, Hans J},
  journal={Physical Review Letters},
  volume={59},
  number={20},
  pages={2326},
  year={1987},
  publisher={APS}
}

@article{herrmann1990damage,
  title={Damage spreading},
  author={Herrmann, HJ},
  journal={Physica A: Statistical Mechanics and its Applications},
  volume={168},
  number={1},
  pages={516--528},
  year={1990},
  publisher={Elsevier}
}

@article{grassberger1995damage,
  title={Are damage spreading transitions generically in the universality class of directed percolation?},
  author={Grassberger, Peter},
  journal={Journal of Statistical Physics},
  volume={79},
  number={1},
  pages={13--23},
  year={1995},
  publisher={Springer}
}

@article{bagnoli1999synchronization,
  title={Synchronization and maximum Lyapunov exponents of cellular automata},
  author={Bagnoli, Franco and Rechtman, Ra{\'u}l},
  journal={Physical Review E},
  volume={59},
  number={2},
  pages={R1307},
  year={1999},
  publisher={APS}
}

@article{Acebron-etal-05,
author = {J. A. Acebr{\'o}n and L. L. Bonilla and C. J. P{\'e}rez Vicente and 
F. Ritort and R. Spigler},
collaboration = {},
title = {The {K}uramoto model: {A} simple paradigm for synchronization phenomena},
publisher = {APS},
year = {2005},
journal = {Rev. Mod. Phys.},
volume = {77},
number = {1},
pages = {137-175}
}

@article{chen2019dynamics,
  title={Dynamics of the Kuramoto-Sakaguchi oscillator network with asymmetric order parameter},
  author={Chen, Bolun and Engelbrecht, Jan R and Mirollo, Renato},
  journal={Chaos: An Interdisciplinary Journal of Nonlinear Science},
  volume={29},
  number={1},
  year={2019},
  publisher={AIP Publishing}
}

@article{brister2020three,
  title={When three is a crowd: {C}haos from clusters of {K}uramoto oscillators with inertia},
  author={Brister, Barrett N and Belykh, Vladimir N and Belykh, Igor V},
  journal={Physical Review E},
  volume={101},
  number={6},
  pages={062206},
  year={2020},
  publisher={APS}
}

@article{maistrenko2005chaotic,
  title={Chaotic attractor in the {K}uramoto model},
  author={Maistrenko, Yuri L and Popovych, Oleksandr V and Tass, Peter A},
  journal={International Journal of Bifurcation and Chaos},
  volume={15},
  number={11},
  pages={3457--3466},
  year={2005},
  publisher={World Scientific}
}

@book{Gupta-Campa-Ruffo-18,
	author = {Gupta, Shamik and Campa, Alessandro and Ruffo, Stefano},
	title = {Statistical Physics of Synchronization},
	publisher = {Springer},
	year = {2018},
	address = {Cham},
}

@article{teramae2007reliability,
  title={Reliability of temporal coding on pulse-coupled networks of oscillators},
  author={Teramae, Jun-nosuke and Fukai, Tomoki},
  journal={arXiv preprint arXiv:0708.0862},
  year={2007}
}

\end{document}